**User Involvement in Robotic Wheelchair Development: A Decade of Limited Progress - A Narrative Review**

Mario Andres Chavarria

Geneva School of Health Science, University of Applied Sciences and Arts Western Switzerland, Geneva, Switzerland

Santiago Price Torrendell

Institute Of Systems and Information Engineering, University of Tsukuba, Tsukuba, Japan

Aude Billard

Learning Algorithms and Systems Laboratory, Ecole Polytechnique Fédérale de Lausanne, Lausanne, Switzerland

Samia Hurst

Institute of Ethics, History, and Humanities, University of Geneva, Geneva, Switzerland

Sébastien Kessler

Grand Conseil, État de Vaud, Lausanne, Switzerland / id-Geo Sàrl, Lausanne, Switzerland

Michael Stein

Harvard Law School Project on Disability, Harvard University, Cambridge, MA, United States

Kenji Suzuki

Institute Of Systems and Information Engineering, University Of Tsukuba, Tsukuba, Japan

Sophie Weerts

Swiss Graduate School of Public Administration, University of Lausanne, Lausanne Switzerland

Diego Paez-Granados*

SCAI-Lab, Department of Health Sciences and Technology (D-HEST), ETH Zurich, GLC, Gloriastrasse 37/39, 8092, Zurich, Switzerland / Swiss Paraplegic Research (SPF), Guido A. Zäch Strasse 4, 6207, Nottwil, Switzerland

Minerva Rivas Velarde*

Geneva School of Health Science, University of Applied Sciences and Arts Western Switzerland, Geneva, Switzerland / Institute of Ethics, History, and Humanities, University of Geneva, Geneva, Switzerland

**Abstract:** Robotic wheelchairs (RWs) offer significant potential to enhance autonomy and participation for persons with mobility impairments, yet many systems fail to achieve sustained real-world adoption. This narrative literature review examined the extent and quality of end-user involvement in RW design, development, and evaluation over the past decade (2015–2025), assessed against core principles shared by major user-involvement design approaches (e.g., user-/human-centred design, participatory/co-design, and inclusive design). Findings indicate that user involvement remains limited and predominantly concentrated in late-stage evaluation rather than early requirement definition or iterative co-design. Of 399 records screened, only 23 studies (≈6%) met the inclusion criteria of verifiable end-user involvement, the and many relied on small samples (often around ten participants ), with limited justification for sample size selection, proxy users, laboratory-based validation, and non-standardized feedback methods. Research teams were largely engineering-dominated (≈89%) and geographically concentrated in high-income countries. Despite strong evidence that sustained user engagement improves usability and adoption in assistive technology, its systematic implementation in RW research remains rare. Advancing the field requires embedding participatory methodologies throughout the design lifecycle and addressing systemic barriers that constrain meaningful user involvement.

**Additional Keywords:** Robotic wheelchairs, assistive technology, user-centered design, participatory design, co-design, human–robot interaction, assistive robotics, user involvement, accessibility, intelligent wheelchairs

## 1 INTRODUCTION

Mobility impairments pose significant challenges to autonomy and social inclusion, directly impacting the quality of life for individuals affected. Robotic wheelchairs (RWs) represent a transformative advancement in mobility solutions for individuals with severe disabilities, offering potential enhancements in autonomy, social integration, and overall quality of life. Defined as powered wheelchairs equipped with automatically controlled, reprogrammable capabilities for sensing, planning, and navigating environments, RWs are a natural evolution from powered wheelchairs (EPWs), which typically provide motorized propulsion under continuous user command via an input device (1). The integration of advanced technologies such as sensors, actuators, and control systems enables RWs to address challenges associated with complex navigation scenarios, facilitating activities of daily living and community participation (1,2,3).

Despite the potential benefits of RWs, their widespread adoption remains limited, pointing to broader systemic challenges that extend beyond technological development to include procurement structures, insurance frameworks, and market accessibility for persons with disabilities (1,4). Consumer surveys consistently highlight unmet needs in wheelchair design, human–machine interfaces, assistive robotics, and intelligent systems, while emphasizing affordability and inclusivity in research and development efforts (1,4). These findings underscore the need to embed robust user-focused design approaches in RW development to ensure that solutions align with the specific needs of individuals with disabilities and prioritize affordability, accessibility, and functional effectiveness. Organizations such as WHO and UNICEF also reinforce this perspective, advocating for the involvement of users and their families throughout the entire access pathway to ensure that technologies address real-world needs (5).

There is a pervasive lack of empirical evidence and methodological rigor across many AT design frameworks, where minimal or superficial user interactions are often portrayed as sufficient. Hence, compliance with HCD principles is often lacking (6–10). Adherence to standards such as ISO 9241-210 (11) is frequently weak, with many projects failing to ensure meaningful user participation throughout the development lifecycle (6,7).

The gap in user-centered development is particularly evident in robotic wheelchair research. Literature reviews spanning from 1990 to the present consistently highlight a scarcity of user-centered RW research and low readiness of these technologies for market adoption (1,12–18). A review of 29 smart wheelchair projects conducted by Desai et al. (14) revealed significant shortcomings in user engagement, with most designs overlooking essential elements such as user comfort, usability, and affordability. Many prototypes were based on add-ons or modified EPWs with rigid hardware and software that did not adequately align with user needs or expectations (14). This disconnect is further reinforced by a mismatch in perceptions: while AT providers often believe that users are actively involved in the design process, the users themselves frequently report little meaningful participation, highlighting a persistent disparity in perspectives (4). Further analysis by Sundaram (13), Leaman (12), Tao (15), and Desai (14) shows that RW research and development often stagnates at autonomy level 2 (partial automation), reflecting the difficulties in advancing RW systems beyond basic functionalities. This limited progress is partly attributed to insufficient end-user engagement, as participatory action design and engineering methods remain underutilized despite RWs being identified as a consumer priority (1,19,20).

Greater implementation of participatory design methodologies and direct collaboration with end-users throughout the development process could help overcome existing barriers, ensuring that RWs better address the diverse needs and expectations of intended users. Such approaches are essential not only for improving user acceptance and functional relevance but also for enhancing the commercial viability, societal impact, and long-term sustainability of RW technologies.

### *1.1 Design Philosophies Focused on User Involvement*

*Since the 1970s, numerous design approaches emphasizing user involvement have emerged. Among the most influential are Participatory or Cooperative Design (PD/CD), in which users act as active partners in shaping technological solutions; User-Centered Design (UCD), which positions users as central informants throughout iterative development processes; Co-Design approaches, Human-Centered Design (HCD), which extend participation to broader stakeholder groups and promote co-creation; and Universal or Inclusive Design, which proactively addresses diversity and accessibility across populations (11,21–25).*

- ***Participatory Design (PD) / Co-design participatory** methodologies in which users and other stakeholders actively collaborate in generating, shaping, and refining solutions throughout the development process.*
- ***User- / Human-Centered Design (UCD/HCD):** design approaches that place users at the center of development, prioritizing their needs, experiences, and contexts throughout iterative design and evaluation processes.*
- ***Heuristic approach:** an experience-based design strategy relying on expert judgment and iterative refinement rather than a formally structured participatory framework.*

*Despite their differences in scope and emphasis, these approaches share core principles that define the essence of user-involvement philosophies: direct end-user and stakeholder participation across design stages; multidisciplinary collaboration; systematic user evaluation throughout development; iterative redesign based on user feedback; mutual learning between designers and users; and contextualization within real-world use environments (11,21–25). These shared principles inform the analytical framework of this review, guiding our examination of how user involvement is embedded across development phases, team composition, evaluation practices, and redesign cycles.*

***Parallels with Participatory Action Research:** Participatory design (PD) and participatory action research (PAR) both emphasize collaboration between researchers, designers, and stakeholders to address real-world problems. Waknic et al. (26) describe PD as an approach in which users participate directly in iterative design processes across multiple stages of development. Their review highlights that many PD studies are case-based and that considerable variation exists in how PD is defined and implemented; however, meaningful stakeholder participation generally contributes to improved design outcomes and greater inclusivity*

*Participatory action research follows a similar collaborative logic, in which participants help identify problems, implement actions, and evaluate outcomes through iterative cycles of reflection and practice (27). Its goal is not only to generate knowledge but also to empower communities and contribute to practical social change.*

*Both approaches therefore position users and communities as active contributors rather than passive subjects of research. Integrating such perspectives into technological development can strengthen attention to issues of social justice (6) and fundamental freedoms that assistive technologies often claim to promote (6,28). Taken together, PD and PAR illustrate a broader shift toward research and design practices that treat communities as partners in knowledge creation and problem-solving.*

### *1.2 Gathering users' experiences*

*User involvement in assistive technology (AT) development should be examined not only in terms of whether participation occurs, but also in relation to the methodological approaches used to capture and interpret user perspectives. Participation alone does not guarantee meaningful integration if the tools used to gather experiences, expectations, and performance data are inconsistently selected, inadequately adapted, or insufficiently reported. The choice and application of data collection instruments directly influence the validity, comparability, and interpretability of findings. This section therefore outlines the principal qualitative, standardized, and performance-based methods commonly used to capture user perspectives, providing a conceptual framework for examining their application in robotic wheelchair research.*

*Qualitative methods such as interviews, focus groups, roundtable discussions, and Q&A sessions allow users to articulate their needs, challenges, and preferences in their own words. These approaches are particularly valuable during requirement elicitation and exploratory evaluation, as they provide insight into context, lived experience, and nuanced usability concerns (29–31). Open-ended questionnaires serve a similar function, generating narrative responses that can later be thematically analyzed (32,33). In contrast, think-aloud protocols require participants to verbalize their thoughts while interacting with a system, enabling researchers to identify usability barriers and understand cognitive processes during task execution (31).*

*In addition to user narratives, expert-based usability inspection methods are also employed. The cognitive walkthrough technique involves experts systematically analyzing task sequences from the perspective of a first-time user in order to detect barriers to learnability and potential breakdowns in task flow (34,35). Similarly, heuristic evaluation assesses interface design against established usability principles, such as Nielsen's heuristics, to systematically detect usability issues (31,36).*

*While qualitative and inspection methods emphasize depth and exploratory insight, standardized usability and satisfaction scales provide structured and comparable measures of user perception. Unlike qualitative approaches, which generate rich descriptive data, standardized instruments produce quantitative scores that facilitate benchmarking across systems and iterations (31,37). The System Usability Scale (SUS) (38) is a widely used 10-item, 5-point Likert-scale questionnaire designed to provide a rapid and reliable measure of overall perceived usability (31,37,39). The USE questionnaire evaluates usefulness, ease of use, and satisfaction across multiple dimensions, while the IBM Computer Usability Satisfaction Questionnaire (CSUQ) (40) focuses on system usefulness, information quality, and interface quality. The Questionnaire for User Interaction Satisfaction (QUIS) offers a more detailed assessment of satisfaction with specific interface components and screen elements (31). Tools specific for Assistive technology such as QUEST (and QUEST 2.0) (41) measure satisfaction with device characteristics and service delivery, and the Wheelchair Skills Test Questionnaire (WST-Q) (31,42) provides a self-reported assessment of perceived wheelchair skill performance. Broader outcome instruments, such as QOLRW, extend evaluation to perceived impact on quality of life (43).*

*Beyond usability and satisfaction, several instruments focus on workload and cognitive load. The NASA Task Load Index (NASA-TLX) (44) is a multidimensional workload scale measuring mental demand, physical demand, temporal demand, performance, effort, and frustration (31,45). The Driving Activity Load Index (DALI) adapts workload assessment to*

*driving contexts, while the Subjective Workload Assessment Technique (SWAT) provides a rapid three-level classification of cognitive load (45). The Subjective Workload Dominance (SWORD) technique differs in that it uses pairwise comparisons to rank tasks relative to each other rather than assigning absolute workload scores (31,46).*
*Finally, some instruments assess broader aspects of acceptance and impact. The Technology Acceptance Model (TAM) measures perceived usefulness and ease of use (37), whereas user experience metrics such as Net Promoter Score (NPS) and Customer Satisfaction Score (CSAT) quantify overall satisfaction and loyalty (37). Adoption-focused tools such as ATD-PA estimate the likelihood of successful assistive technology uptake, and CATOM evaluates the impact of assistive devices on caregivers (47). WHODAS 2.0 provides a standardized assessment of disability and functioning, offering contextual insight into users' capabilities (48).*

Given methodological weakness identified in robotics literature, this paper aims to examine how effectively users have been involved in the development of robotic wheelchairs over the past decade, assessed against the core principles shared by major user-involvement design philosophies. Specifically, it seeks to identify the methods and tools employed to engage users throughout the design and evaluation process, evaluate the depth and quality of their involvement, and highlight opportunities for strengthening participatory practices in future RW development. By synthesizing recent advancements and recurring limitations, this review contributes to a deeper understanding of how user engagement can shape more accessible, acceptable, and impactful robotic mobility solutions.
The structure of this paper is as follows: Section 2 describes the methods used to conduct this narrative literature review, including the search strategy, eligibility criteria, and analytical framework. Section 3 presents the results, organized according to participant characteristics, phases of involvement, team composition, evaluation practices, and redesign processes. Section 4 discusses the systemic patterns emerging from these findings and their implications for participatory design in robotic wheelchair development. Finally, the conclusion outlines recommendations and directions for future research.

## 2 METHODS

### 2.1 Search strategy and scope

Given the narrative nature of this review, we aimed to identify and synthesize the most relevant literature addressing user involvement in the design, development, and evaluation of robotic wheelchairs over the past decade. The review was guided by the following research question:
*How effectively have users been involved in the development of robotic wheelchairs over the past 10 years, and what methods and tools have been used to involve users throughout the design and evaluation process?*
To address these questions, a literature search was conducted across the following databases: IEEE Xplore, ACM Digital Library, PubMed, Web of Science, and Google Scholar. The search covered publications from January 2015 to December 2025 and was limited to articles written in English, French, or Spanish. Search equations were adapted to the syntax of each database, while maintaining the same keywords and Boolean logic. The core search terms combined concepts related to robotic wheelchairs, user involvement, and user-centered design approaches, using the following structure: ("robotic wheelchair*" OR "smart wheelchair*") AND ("user*" OR "patient*" OR "person with disabilit*" OR "participant*") AND ("user-centered design" OR "user-centred design" OR "human-centered design" OR "human-centred design" OR "participatory design" OR "co-design" OR "codesign" OR "inclusive design" OR "user* involvement" OR "user* participation").

### 2.2 Eligibility criteria

Studies were selected based on their relevance to the research question and the quality and clarity of their reporting.
**Inclusion criteria** were:
(i) peer-reviewed journal articles and conference papers;
(ii) Publications between 2015 and 2025; this cut point was chosen because AI breakthroughs around 2015 marked a major turning point for robotics, particularly in perception, natural language, and signal processing (49–51).
(iii) articles written in English, French, or Spanish;
(iv) studies describing the design, development, or evaluation of robotic wheelchairs;
(v) explicit involvement or interaction with end users; and
(vi) experimental, qualitative, mixed-methods, or design case studies focusing on user involvement.
**Exclusion criteria** included:
(i) studies without any form of end-user involvement;
(ii) studies focusing exclusively on manual or powered (non-robotic) wheelchairs;
(iii) reviews, commentaries, patents, or purely theoretical papers;
(iv) studies relying solely on simulations or proxies (e.g., dummies or exclusively able-bodied participants without disabilities); and
(v) publications prior to 2015 or with inaccessible full texts.

### 2.3 Study selection and data extraction

The database search yielded 462 records, of which 399 remained after duplicate removal. An initial screening based on titles, abstracts, and a preliminary review of the full text resulted in 50 papers selected for full-text analysis. This first screening focused on identifying studies that presented the design and development of robotic wheelchairs and included some form of user involvement, while excluding irrelevant publication types and non-robotic mobility devices. To ensure conceptual clarity, we applied a technical inclusion criterion defining "robotic/intelligent" wheelchairs as powered wheelchairs that incorporate on-board sensing and computational control often combined with dedicated electromechanical hardware upgrades, to provide automated or assisted functions beyond direct user actuation. In practice, this included systems implementing shared or autonomous assistance during mobility tasks (e.g., navigation support or collision avoidance, curb/step climbing), intent-aware interaction beyond a conventional joystick, and/or adaptive control that modifies wheelchair behavior in response to the user or environment.

A second, in-depth screening of the full texts was then conducted to assess the nature, clarity, and rigor of reported user involvement. During this stage, studies that claimed a user-centered design approach but failed to report concrete user participation were excluded (this will be discussed further in **Section 3.1**). Studies involving end users as well as mixed samples (i.e., end users together with proxies) were retained. Following this process, 23 studies were included in the final narrative synthesis: (29,30,32,33,39,52–69)

Each article was independently analyzed by at least two authors, who evaluated titles, abstracts, full texts, and scientific validity (70). Any discrepancies in inclusion decisions were resolved through discussion until consensus was reached. References cited within the selected articles were also screened to identify additional relevant studies.

In addition, the core principles derived from user-involvement design philosophies (**Section 1.1**) were operationalized as analytical categories guiding data extraction and synthesis. Specifically, we assessed whether each study demonstrated: (i) early and continuous user participation, (ii) multidisciplinary collaboration, (iii) systematic user evaluation practices, (iv) iterative redesign based on user feedback, and (v) contextualization within real-world environments. These dimensions formed the analytical framework through which the depth and quality of user involvement were evaluated across the included studies.

Several additional papers identified during screening claimed user-centred design, human-centred design, participatory design, co-design, or similar approaches but were excluded during the second filtering stage due to the absence of verifiable end-user involvement. These studies either did not report any user participation (71–73) or relied on proxies, such as virtual users, simulated datasets, able-bodied participants, dummies, literature-based assumptions, or framed "user involvement" as the inclusion of customizable system features rather than direct participation (45,74–86). Others mentioned user involvement only briefly, without specifying who was involved, how many participants were included, or the nature of their contribution (85,87,88). In total, 27 of the 50 papers that passed the initial screening were excluded at this stage, indicating that a substantial proportion of studies claiming participatory approaches did not provide clear evidence of direct engagement with end users with disabilities.

# 3 RESULTS

The results indicate that user engagement in robotic wheelchair research remains largely consultative and post hoc, with limited evidence of sustained, structurally embedded participatory design shaping core technological architectures. This section synthesizes evidence from 23 publications (corresponding to 13 distinct research projects) examining user involvement in the development of robotic wheelchairs. The findings are divided into three sections, presented as follows: a) focusing on who was involved, b) when and how participation occurred, and c) whether it substantively influenced design outcomes.

The 23 included papers (from 399 unique records screened) corresponded to 13 distinct research projects. Most of the identified research is concentrated within two major international initiatives: the ADAPT project in Europe (30,62,64–67) and the MEBot project at the Human Engineering Research Laboratories (HERL) in the United States (39,52,55,63,69), which together account for nearly half of the included literature.

## 3.1 Who was involved?

The total number of participants across the individual studies varied considerably, ranging from a minimum of one participant, reported in (53,60,61), to 191 individuals in (29), which included 186 valid survey respondents and five on-site patients. Across the reviewed literature, the most common sample size was approximately 10 to 11 participants (39,52,55,56,69). None of the papers included provided a formal explanation of how sample sizes were calculated. However, several authors offered justifications for the relatively small samples by pointing to challenges associated with recruiting larger cohorts. For instance, Ezeh et al. (64) noted that "it was difficult recruiting large numbers of wheelchair users for all our experiments." Similarly, Candiotti et al. (55) reported that their "sample size was limited to ten participants due to the limited population of EPW users where the study was conducted," and attempted to justify this limitation by citing a work by Virzi (89), who suggested that "80% of usability feedback can be detected with four or five participants." However, in a subsequent study (39) the same research group acknowledged that the small sample size "may have limited the statistical power of our analyses and ability to avoid Type 2 (false negative) errors with respect to user perceptions." Finally, Leblong et al. (66) also addressed this issue, arguing that, despite involving what might appear to be a small group (23 wheelchair users), their sample size was "relatively high in view of the data existing in the literature" for smart wheelchair trials.

The research demonstrates a focus on specific chronic conditions, with spinal cord injury (SCI) being the most frequently reported disability across the literature, followed by cerebral palsy (CP), multiple sclerosis (MS), stroke, and acquired brain injury. Four studies (29,33,57,58) did not report clinical diagnoses, referring to participants only as "patients," "manual wheelchair users," or "persons with disability" without further specification.

**Mixed Groups and Proxy Use:** Approximately one third of the analyzed works employed mixed cohorts composed of both end users with disabilities and able-bodied proxies to evaluate system performance. Proxies were commonly used as a baseline for comparison or when recruitment of wheelchair users was logistically challenging. Authors reported several motivations for this approach. Recruitment difficulties were frequently cited; for example, Ezeh et al. (64) explained that they included able-bodied participants because "it was difficult recruiting large numbers of wheelchair users for all our experiments". Other studies used healthy participants as novice comparison groups alongside experienced wheelchair users, including Cruz et al. (54) (7 able-bodied proxies / 6 end users), Ezeh et al. (64) (4 able-bodied proxies / 3 end users), Delmas et al. (65) (17 able-bodied proxies / 2 end users), and Morbidi et al. (30) (17 able-bodied proxies / 102 end users). In requirement-identification phases, proxies were sometimes involved as stakeholders or domain experts rather than drivers, as in Gomez Torres et al. (32) (52 end users/ 18 stakeholders) and Leaman et al. (60) (1 end user/ multiple able-bodied students). Finally, Priyanayana et al. (57) (25 end users/ 6 'healthy caretakers') incorporated proxies in a comparative "control experiment" contrasting intelligent wheelchair use with manual assistance provided by a caregiver.

Table 1: Summary of participant characteristics and reporting across included studies

| **Dimension** | **Key Findings** | **Reporting Frequency** |
|---|---|---|
| **Sample size** | Range: 1 to 191 participants; most common ≈10–11 | All studies reported N; none reported formal sample size calculation |
| **Justification of sample size** | Recruitment challenges; usability-based arguments; occasional acknowledgment of limited statistical power | No formal power analyses reported |
| **Disability/health status** | Predominantly SCI; also CP, MS, stroke, acquired brain injury; some unspecified | 18/23 (78%) |
| **Wheelchair experience** | Strong reliance on highly experienced users (≈16+ years); some studies did not specify experience | 18/23 (78%) |
| **Gender** | Frequently male-dominated (≈2:1 or higher); many studies did not report gender | 13/23 (57%) |
| **Age** | Range: 14–76+ years; weighted mean ≈48.9 years; limited inclusion of elderly users | 11/23 (48%) |
| **Weight** | Reported mainly as safety/mechanical constraint (39–113.4 kg); sometimes simulated users | 7/23 (30%) |
| **Inclusion/exclusion criteria** | Explicitly defined criteria | 5/23 (22%) |
| **Use of proxies** | ≈1/3 used mixed cohorts (end users + able-bodied participants or caregivers) | ~33% |

### 3.2 Stakeholder Participation

End-user involvement is a core principle of user-centered design (UCD), but it is not sufficient on its own. UCD also emphasizes multidisciplinary teams, the inclusion of users as co-designers, stakeholder participation, and sensitivity to diverse contexts of use. This section examines how these dimensions are addressed in robotic wheelchair research. A detailed summary of the data discussed in this section is provided in Appendix A.1.

Across the analyzed studies, research teams were overwhelmingly dominated by engineering profiles. On average, approximately 89% of team members had an engineering background, indicating a strong technical orientation in robotic wheelchair R&D. In 14 out of 23 studies, teams were composed exclusively of engineering or technical disciplines, with no explicit involvement of health, social science, or design professionals (29,33,53,56–65,68).

A smaller subset of studies demonstrated moderate multidisciplinarity (30,39,52,55,63,66,67,69), typically through the inclusion of healthcare professionals alongside engineers. Only three studies explicitly involved sociologists (30,66,67), one included a clinical psychologist (54), and one involved a product designer (32).

#### *3.2.1 Presence of Persons with Disabilities Within Development Teams (Co-Design)*

Direct participation of persons with disabilities (PWD) as members of development teams was rarely and inconsistently reported. Most studies described user involvement primarily during testing or evaluation phases, rather than through sustained participation as co-researchers or co-designers. In a limited number of projects, most notably the MEBot series (39,52,55,63,69) and Leaman et al. (60), collaboration with rehabilitation professionals and long-term wheelchair users approached co-design practices. However, explicit identification of authors or team members as PWD remained uncommon.

This limited reporting may reflect both underrepresentation and under-documentation. While user testing was widespread across the analyzed studies, the integration of PWD as equal partners within research teams, a central principle of co-design frameworks, was comparatively rare. This finding highlights a persistent gap between user participation and genuine participatory development in robotic wheelchair research.

#### *3.2.2 Contextual Diversity and Multi-Context Development*

Context analysis reveals a persistent imbalance in robotic wheelchair research. Development is predominantly shaped by high-income country (HIC) perspectives, with limited engagement across diverse economic, infrastructural, and cultural settings. Although some LMIC-based studies address affordability and contextual constraints, systematic multi-context co-development remains rare.

Approximately 65% to 70% of the analyzed studies were led by research centers based in HICs, such as the USA, France, the UK, and Japan (30,32,39,52–55,60–67,69). These studies typically reflect research priorities, infrastructures, and regulatory environments characteristic of well-resourced healthcare and research systems.

In contrast, studies originating from low- and middle-income countries (LMICs), India, Sri Lanka, the Philippines, Mexico, and Indonesia, were fewer and predominantly focused on localized development efforts. These projects generally addressed affordability, robustness, and contextual constraints by adapting or augmenting basic manual or low-cost powered wheelchairs, rather than developing highly automated or sensor-rich robotic platforms (29,33,56–59,68). While sensitive to local needs and resource limitations, their scope was typically confined to national or regional contexts.

Only one study explicitly adopted a global, multi-context perspective. Gomez Torres et al. (32) bridged development in a HIC setting (USA) with user input from seven countries spanning both HIC and LMIC contexts (India, Sri Lanka, Egypt, Mexico, UAE, UK, and USA), explicitly acknowledging contextual diversity in user needs, infrastructure, and cultural expectations.

### 3.3 When and how did participation occur?

Table 2 summarizes how user and stakeholder involvement is reported across the reviewed studies, detailing for each project the number and type of participants (end users versus proxies), as well as the stages of the design process in which they were involved. The table highlights that the majority of studies primarily engaged participants during evaluation and testing phases (i.e., laboratory trials, clinical or field assessments, competitions), frequently combining performance metrics with subjective measures such as workload, satisfaction, or usability questionnaires (e.g., NASA-TLX, SUS).

Table 2: Summary of User and Stakeholder Involvement Across Reviewed Studies, Including Participant Type, Sample Size, Design Phases, and Nature of Engagement

| Ref | Involved population of users | Phases with Participant Involvement | Nature of Involvement |
|---|---|---|---|
| (29) | 191 end users + non-defined test "subjects" | Requirements and Evaluation | Questionnaires and expert consultation for requirements; EEG recording of subjects to validate BCI system accuracy. |
| (52) | 11 end users | Evaluation/ Testing | 11 participants performed 18 trials of curb negotiation to evaluate efficiency and effectiveness based on ISO standards. |
| (53) | 1 pilot | Evaluation/ Testing | A selected pilot participated in the 2016 Cybathlon competition to provide performance and stability metrics. |
| (54) | 13 participants: 6 end users + 7 Proxies able-bodied | Evaluation/ Testing | Navigation experiments in office settings; users provided subjective feedback via NASA-TLX and satisfaction surveys. |
| (32) | 70 participants: 52 End Users + Proxies: healthcare professionals, and non-profits. | Requirements and Design (feedback) | Stakeholders across 7 countries provided input via open-ended surveys and interviews to create design character profiles (requested feedback throughout the design process). |
| (30) | 119 participants: 102 End Users + Proxies:17 able-bodied participants. | Requirements and Evaluation | Roundtable sessions for requirement definition; users performed clinical trials using standardized courses. |
| (62) | 23 end users | Evaluation/ Testing | Randomized double-blind clinical trials in hospital hallways; satisfaction surveys. |
| (66) | 23 end users | Evaluation | Single-blind trials on standardized circuits of increasing difficulty; cognitive load and usability assessment. |
| (67) | 16 end users | Evaluation/Validation | Randomized trials comparing assisted vs. non-assisted driving; user feedback. |
| (65) | 19 participants: 2 End Users + Proxies:17 able-bodied participants. | Evaluation/ Testing | Users navigated challenging obstacle courses (elevators/slaloms) and provided viewpoint preferences in written surveys. |
| (55) | 10 End Users | Evaluation and feedback (Requirements) | 11 users tested an early version with sequential steps and provided feedback to improve a new prototype; 10 users performed trials to rate workload and usability. |
| (56) | 10 "individuals" | Evaluation/ Testing | Quantitative testing where participants were measured during sitting-to-standing robotic transitions. |
| (57) | 31 participants: 25 End Users + Proxies: 6 caretakers | Evaluation/ Testing | Users provided voice commands to compare robotic execution against human caretaker perception in control experiments. |
| (58) | 30 Persons with Disability | Evaluation/ Testing | 30 participants performed final acceptance testing and rated the prototype on a Likert scale for ease of use. |
| (59) | 10 "Volunteers" | Evaluation/ Testing | Volunteers verified the accuracy and viability of four different control modalities in controlled lab settings. |
| (60) | 1 end user (Co-author) Proxies: students | Requirements, Co-Design, and Evaluation | Continuous Q&A sessions for design utility (1 author is a wheelchair user); virtual simulations and initial human trials with qualitative interviews. |
| (39) | 10 end users | Requirements and Evaluation | Feedback from previous evaluations used to optimize the technology; users tested dynamic suspensions on ramps and potholes. |
| (61) | 1 driver | Evaluation/ Testing | Kinematic mobility performance analysis comparing simulations against experimental data from a driver. |
| (63) | 43 end users | Requirements and Evaluation | 31 users in focus groups identified barriers; 12 new users provided feedback on CAD models via likelihood-of-use surveys. |
| (69) | 11 participants:10 end users + 1 Proxy: therapist | Evaluation | Lab testing of 33 tasks to measure physical and mental workload. |
| (64) | 7 participants: 4 Proxies: able-bodied novices + 3 wheelchair users | Evaluation/ Testing | Comparison trials of shared control approaches; users provided subjective feedback via IBM usability questionnaires. |
| (68) | 50 families with elderly people + 1 proxy: dummy | Requirements | 50 families of elderly people prioritized 11 functional indicators using the Analytic Hierarchy Process (AHP). |
| (33) | 30 participants: Wheelchair users and caregivers. | Requirements | Users and caregivers provided open-ended elicitations and Likert ratings to identify critical customer attributes. |

Analysis of the 23 included studies reveals a clear pattern: user involvement is heavily concentrated in the evaluation and testing phases, with substantially fewer studies involving users during requirement definition and very limited evidence of genuine co-design.

Evaluation and testing: Twenty of the 23 analyzed studies involved participants in the evaluation of their designs (29,30,39,52–67,69). Moreover, evaluation-only involvement represents the dominant model of participation (15/23 studies) (52–59,61,62,64–67,69). In these studies, users primarily evaluated performance, safety, workload, or usability outcomes of existing designs and prototypes.

Requirement definition: Only 8/23 studies involved users during early requirement-definition phases: (29,30,39,60,63) combined requirement elicitation with later evaluation phases, while (32,33,68) involved users exclusively in defining functional priorities, without subsequent evaluation phases. Early participation was most often structured through surveys, focus groups, or interviews rather than collaborative design activities, with decision-making authority primarily retained within engineering teams.

Design phase: Documented co-design, understood as sustained and collaborative involvement of users in shaping core design decisions, was rare. Leaman et al. (60) stands out as the only study explicitly framing the technology as "designed by a scientist with quadriplegia" (J.F. Leaman), who actively defined features and collaborated with students to determine design utility. This represents a clear integration of experiential knowledge into the technological architecture.

Gomez Torres et al. (32) reported maintaining contact with users throughout the design process after initial surveys and interviews, requesting feedback during development. While this reflects ongoing interaction and is more iterative than most studies, it remains unclear to what extent users held decision-making power or whether their role was primarily consultative. Similarly, the ADAPT consortium (30) referenced co-design principles but described user input mainly as contributing to adjustments intended to "improve experience" (e.g., sensor positioning), rather than influencing fundamental mechanical or algorithmic structures.

Other works that can be considered as co-design, since they include prominent authors who are wheelchair users, are (39,52,55,63,69), as discussed in Section 3.2.1. However, their methodologies frame involvement around recruited "participants" or "subjects," without explicitly documenting the authors' lived experience as a formal co-design contribution. While their experiential knowledge likely informed development, it is not methodologically articulated as co-design, possibly reflecting reporting conventions in clinical and engineering research.

### 3.4 Methods and Tools for Participation

This section describes how user involvement was operationalized, highlighting the structured approaches that shaped participation. A detailed summary of the data discussed in this section is provided in Appendix A.2.

#### *3.4.1 Design Philosophies Focused on Users*

Only 10 of the 23 studies explicitly mentioned the implementation of a user-focused design methodology. The reported approaches included Participatory Design (PD) / Co-design (30,32,60,63), User- / Human-Centered Design (UCD/HCD) (29,30,33,39,55,56), Heuristic approach (52). Some works further operationalize these design philosophies through structured models and engineering frameworks. For example, Gomez Torres et al. (32) combine Participatory Design with the Double Diamond (DD) model, which structures innovation into divergent and convergent phases (discover, define, develop, deliver) to guide iterative problem framing and solution development. Similarly, Lukodono et al. (33) combine UCD with a customized engineering framework, AHOQ (Axiomatic House of Quality), which integrates House of Quality and Axiomatic Design to translate user-prioritized requirements into technically decoupled design solutions. Morbidi et al. (30) further propose integrating proxemics principles to account for social navigation and interpersonal distance; however, this approach has not yet been operationalized within the reported framework. Notably, none of the reviewed studies explicitly mentioned adherence to ISO 9241-210: Human-centred design for interactive systems.

### 3.4.2 Data Collection Methods

User-self-reported data collection methods aim to capture users' subjective experiences, perceptions, satisfaction, and cognitive demands when using interactive systems, documenting dimensions of interaction that cannot be inferred solely from performance metrics or technical testing.

Among the 23 included studies, the majority (18/23) reported the use of at least one user-reported data collection method while five studies did not report any user-reported instruments (52,53,56,59,61). Overall, customized questionnaires were the most frequently used approach, appearing in 8 of the 23 studies (32,54,57,58,63,65,66,69). This suggests a strong preference for ad hoc survey instruments tailored to specific prototypes or experimental contexts. While customization allows researchers to target specific design features, it raises methodological questions regarding comparability, validation, and reliability, particularly when existing standardized instruments are available.

Among standardized instruments, the NASA Task Load Index (NASA-TLX) was the most commonly adopted tool, used in 5 studies (54,55,66,67,69) (3 projects: RobChair, MEBot, and ADAPT). This predominance aligns with the broader wheelchair R&D literature, where NASA-TLX is frequently reported as a widely used multidimensional workload assessment tool (45). In contrast, the frequency of other standardized usability and satisfaction scales, such as SUS (2 studies (39,55) / 1 project (MEBot)), USE (3 studies (30,66,67) / 1 project (ADAPT)), QUEST/QUEST 2.0 (2 studies (39,62)), and WST-Q (2 studies (66,67) /1 project (ADAPT)), CSUQ (1 study (64)), and ENASE-based Questionnaires (1 study (33)), was considerably lower, indicating limited convergence around a shared set of standardized measures.

Qualitative methods were moderately represented. Interviews were used in 4 studies (29,32,68,69), focus groups in 2 (63,69), and open-ended questionnaires in 2 (32,33), while roundtables (30) and Q&A sessions (60) appeared in only one study each. This suggests that exploratory qualitative inquiry was present but not dominant across the corpus.

Several studies combined multiple user-reported methods, reflecting a triangulated approach to subjective evaluation. These multi-method designs suggest an emerging tendency toward triangulation of subjective metrics, integrating qualitative insights with standardized usability and workload scales to provide a more comprehensive account of user experience.

A summary of the identified user-reported data collection methods organized by category are presented in Appendix A.3.

### 3.4.3 Experimental and Observational Data Collection Methods

Most of the analysed studies (16/23) relied primarily on interaction and performance metrics (task-based measures) to evaluate robotic wheelchair systems (30,39,52–57,59,61,62,64–67,69). These measures typically quantified driving performance, collision frequency, navigation efficiency, task completion time, distance travelled, stability, or transition times. Several studies adopted comparative designs to assess system improvements, including assisted versus non-assisted modes (67), intelligent versus manual wheelchair control (57), and simulated versus real-world testing conditions (61).

Only four studies incorporated standardized clinical performance assessment tools, which provide structured and validated evaluation protocols: the Wheelchair Skills Test (WST), used to assess functional driving skills (64,66,67); the Power Mobility Indoor Driving Assessment (PIDA), used to evaluate indoor powered wheelchair navigation performance (66,67); and the Powered Mobility Clinical Driving Assessment (PMCDA), used to assess clinical driving ability (69).

Controlled experimental protocols were explicitly reported in only three studies. These included random double-blind clinical trials (62), single-blind randomized trials (66), and randomized comparisons of assisted versus non-assisted modes (67), strengthening methodological rigor through controlled study design.

Beyond task performance metrics, Zhang et al. (29) incorporated physiological measurements (EEG recordings) to assess brain–computer interface accuracy, representing the only study employing biometric data. Simulation-based approaches were reported in four studies: virtual simulations (60), Open Dynamics Engine simulations (61), Finite Element Analysis (33), and CATIA ergonomic simulations (29). Wang & Wang (68) conducted technical testing using dummies rather than real users. Three studies (32,58,63) did not report experimental or observational performance-based data collection methods.

Notably, experimental interaction protocols commonly used in technology development, including assistive technologies, such as the Wizard of Oz method (90,91), where system autonomy is simulated during early-stage testing, and serious games involving gamified, structured task-based evaluations/training (92,93), were not observed in the reviewed sample.

*3.4.4 Testing Environments*

Across the studies, testing was conducted predominantly in laboratory-controlled or simulated environments, even when aiming to replicate real-world features. Most studies (15/23) conducted testing in controlled laboratory settings (indoor only: 8 studies (30,56,57,59,63,65–67); outdoor only: 6 studies (39,52,55,61,64,69); indoor & outdoor: 1 study (53)), prioritizing standardized obstacles such as ramps, curbs, corridors, and elevators. Simulations appeared in 6 studies, including virtual environments (60), CATIA ergonomic simulations (29), Open Dynamics Engine simulations (61), and Finite Element Analysis (33). Only two studies conducted testing under real-world conditions: Devigne et al. (62) in the hallway of a rehabilitation center during working hours with dynamic pedestrian traffic, and Cruz et al. (54) in a real indoor office environment with hallways, narrow doorways, and office obstacles.

However, direct user consultation in defining test protocols was rare. Only Daveler et al. (63) explicitly based test conditions on feedback from experienced users, identifying 23 driving conditions through user consultation. Three studies reported partial involvement: Candiotti et al. (55) optimized strategies based on previous user feedback; Leaman et al. (60) incorporated a co-author end-user perspective in framework design; and Candiotti et al. (39) simulated tasks reflecting real-world situations reported in prior literature on Veteran interviews. The remaining 19 studies (83%) defined tests primarily through standards and researcher/expert decisions without direct user input, which may limit ecological validity.

### 3.5 Did User Involvement Influence Design Outcomes?

To assess whether user involvement influenced design outcomes, we examined two criteria: (1) whether users participated in requirement or needs definition prior to system development, and (2) whether prototypes were iteratively modified based on user feedback after testing.

*3.5.1 User Involvement in Requirement Definition*

As reported in Section 3.3 only a subset of studies (≈25%) involved users at the requirement-definition stage in a way that could meaningfully influence initial design decisions. However, the depth and scope of user influence varied substantially across programs. A comparatively strong form of requirement-phase involvement was observed within the MEBot program, where early participatory input and literature reporting interviews of 1022 Veteran wheelchair users informed system priorities (39,63). In contrast, within the ADAPT program, requirement definition relied primarily on clinician and professional consultation (30,66,67), rather than direct user-driven framing of system architecture. While users participated in evaluation phases, the foundational design requirements were largely defined by medical and engineering experts.

*3.5.2 Iterations and redesigns*

Evidence of iterative redesign explicitly driven by user feedback was limited but present.

Within the MEBot program, iterative redesign based on user feedback is clearly documented. Across successive publications, user testing led to structural and algorithmic improvements, including modifications to control interfaces, automation of obstacle negotiation sequences, and redesign of the suspension system to reduce cognitive demand, discomfort, and instability (39,52,55,69). These iterations reflect tangible changes to both system architecture and operational logic following user evaluation.

In contrast, within the ADAPT program, reported iterations were generally limited to ergonomic or parameter adjustments rather than structural redesign. Modifications included changes such as sensor positioning, acceleration smoothing, and interface display preferences (30,62,64,65). Moreover, some ADAPT studies reported evaluation outcomes without describing subsequent redesign phases based on user feedback (66,67).

Although Gomez Torres et al. (32) do not report post-testing redesign iterations, they implemented a participatory design approach characterized by continuous user input and recurrent lab visits by multiple wheelchair users and researchers. Their work reflects iterative consultation during development, even if formal evaluation-based redesign cycles were not documented.

Several studies (33,53,59–61) reported iterative improvements or represent refinement iterations of earlier prototypes. However, these cannot be clearly characterized as user-centered redesign iterations, as the modifications were primarily driven by technical evaluation rather than documented user feedback, or the influence of user input was not explicitly reported. Five studies did not report any redesign iterations (54,56–58,68).

Moreover, 7 of the 23 studies (≈30%) (53,54,56–59,61) did not report any end-user involvement in requirement definition prior to technology development, nor any redesign iterations based on user feedback, nor explicit co-design practices. In these cases, technologies were developed primarily through engineering criteria, with limited evidence that end users influenced the core design trajectory. This pattern underscores a substantial gap between the rhetoric of UCD and the actual degree of user impact in shaping the systems intended for their use.

## 4 DISCUSSION

This narrative literature review examined how effectively users have been involved in the design, development, and evaluation of robotic wheelchairs over the past decade, assessed against the core principles shared by major user-involvement design philosophies, including participatory design, user-centered design, human-centered design, and co-design. Specifically, we analyzed who was involved, when and how they were engaged, and to what extent they contributed to requirement definition, development processes, evaluation practices, and iterative redesign cycles.

The findings indicate that the principles of user-focused design methodologies are not systematically implemented in the development of robotic wheelchairs. Although user involvement is frequently mentioned, it is often limited in scope. The substantial number of studies excluded during the screening process, many of which claimed user-centered or participatory approaches without demonstrating verifiable engagement of persons with disabilities, already highlights the relative rarity of rigorously documented user involvement in this field.

Even among the 23 selected articles that met the inclusion criteria, the key principles that define user-involvement philosophies (early and continuous participation, multidisciplinary collaboration, iterative redesign based on feedback, contextual sensitivity, and genuine co-design) were inconsistently reflected. In most cases, user participation was consultative rather than collaborative, and confirmatory rather than formative.

Across the reviewed studies, stronger and more sustained forms of user involvement are concentrated within a few research programs. The MEBot program (39,52,55,63,69) stands out as the strongest example of sustained user-centered development. It combines early requirement grounding, transparent participant characterization, standardized clinical and usability measures, and documented iterative redesign affecting core mechanical and control architecture. User feedback demonstrably shaped successive prototype generations. The ADAPT program (30,62,64–67) demonstrates methodological rigor, multidisciplinary collaboration, and strong clinical validation using standardized tools. However, user participation is concentrated mainly in evaluation phases, with requirement definition largely expert-driven and redesigns focused more on ergonomic and parameter adjustments than structural changes.

Outside these programs, individual studies demonstrate strengths in specific dimensions, such as participatory framing and multi-country engagement (32), multidisciplinary balance (66,67), or real-world testing (54,62), but these strengths remain fragmented. Overall, user involvement across the field remains predominantly evaluation-based rather than structurally embedded in early design and architectural decision-making. This pattern is not only methodological but also systematic, reflecting broader challenges in how robotic wheelchair technologies are conceived, validated, and translated.

Beyond individual study designs, the review exposes recurring structural patterns that shape the role of users in robotic wheelchair research:

## 4.1 Limited End-User Involvement, Small Samples, and the Use of Proxies

Less than 6% (23/399) of the retrieved papers met the criterion of reporting verifiable end-user involvement. Unfortunately, this finding is not unique to robotic wheelchair research. Comparable reviews in other areas of assistive technology (AT) development report similarly low levels of meaningful user participation, ranging from approximately 11% to less than 1% (6,94,95).

Even when users were included, most studies relied on very small cohorts, often around ten participants, without formal justification of sample size calculations. Recruitment difficulties were frequently cited to explain these small numbers and, in some cases (64,96), to justify the substitution of persons with disabilities with proxies, such as able-bodied participants, students, simulated users, or technical “dummies.”

The combined reliance on small samples and proxy participants raises both methodological and ethical concerns. Although qualitative approaches can justify smaller sample sizes because saturation differs from statistical representativeness, the reviewed studies rarely provided a clear methodological rationale for their sampling or design choices. In studies using quantitative metrics, small samples reduce statistical power and limit generalizability. It is particularly problematic to assume that proxy participants can replace real users while still producing valid and generalizable findings. Such substitution may exclude people with disabilities from providing feedback on technologies intended for their use. Consequently, it can distort usability, workload, and performance findings, particularly when participants do not share the lived experiences and long-term priorities of wheelchair users. While simulation of impairment may be justified in certain circumstances, if not rigorously designed and critically contextualized, it risks oversimplifying and misrepresenting lived experience. This practice has been coined as “Disabled for a Day trap”, which refers to how short-term simulations by non-disabled individuals tend to emphasize temporary discomfort or social perception while overlooking critical long-term concerns such as autonomy, safety, reliability, and the risk of being stranded (97).

Although recruitment barriers are mentioned in the reviewed studies, their underlying causes are rarely examined in depth. Broader literature on user-centered design in assistive technology highlights multiple institutional contributors, including limited time and funding to sustain iterative engagement, academic incentive systems that prioritize technical novelty over participatory processes, regulatory frameworks that are poorly adapted to iterative prototyping (91,98), and paternalistic overprotection of persons with disabilities by investigators, clinicians, or ethics committees (Stineman & Musick, 2001). Ethics procedures often require finalized protocols before any user interaction, which restricts early-stage feedback and may delay or discourage meaningful engagement (91,98).

Whether these patterns stem from lack of prioritization, institutional disincentives, regulatory rigidity, or overprotection framed as safeguarding "vulnerable populations," the empirical picture is consistent: end-user involvement remains rare and, when present, is typically marginal in the development of technologies intended for their use, including robotic wheelchairs.

### 4.2 Fragmented Methods for Capturing User Perspectives

Beyond where and how testing occurs, a further cross-cutting limitation concerns how user perspectives are captured throughout the development process. Across the reviewed studies, both requirement elicitation and post-design evaluation frequently relied on customized questionnaires rather than standardized, validated instruments. While tailored tools may address context-specific features, their widespread use raises concerns regarding reliability and construct validity and limits comparability across studies. Despite the availability of validated instruments, robotic wheelchair research lacks a consistently adopted methodological framework for capturing user needs and experiences, both during requirement elicitation and post-design evaluation, that would support meaningful cross-study comparison and cumulative knowledge building.

The reasons behind this preference for ad hoc instruments remain unclear. Existing tools may be perceived as insufficiently adapted to robotic wheelchair contexts, or researchers may be unfamiliar with validated instruments outside their disciplinary domain. However, repeatedly developing new, non-validated questionnaires requires time and resources that could instead support participatory refinement within established methodological frameworks. Ultimately, fragmented and non-standardized measurement practices weaken the evidentiary basis for design decisions and constrain the capacity of user input to meaningfully influence core architectural development.

### 4.3 Limited Transparency in User Reporting

Even when end users are involved, their participation is frequently insufficiently documented. Across the reviewed studies, key demographic and methodological details, such as age, gender, weight, disability profile, inclusion and exclusion criteria, and the precise nature of user involvement, were often partially reported or entirely absent. In some cases, it was not even clear whether participants were end users with disabilities or proxies (e.g. (61,68)). This lack of transparency limits reproducibility, reduces interpretability, and complicates the assessment of representativeness and external validity. It is striking that, despite the acknowledged challenges of recruiting and involving end users, their characteristics and contributions are not described with the same level of detail routinely devoted to technological specifications. This imbalance may reflect disciplinary priorities, journal conventions, or a tendency to privilege technical novelty over participatory processes. Regardless of the cause, insufficient reporting of user involvement and if involved how this occurred undermines accountability and reinforces the marginal status of users within the development narrative, highlighting the need for clearer reporting standards in robotic wheelchair research.

### 4.4 Recruitment Bias: Middle-Aged, Male, Veteran-Dominated Samples

Among the studies that did report participant characteristics, a clear demographic pattern emerges: samples were predominantly composed of middle-aged men, frequently recruited from Veteran populations and spinal cord injury (SCI) registries. Authors often justified this distribution by noting the higher prevalence of traumatic SCI among men and the perception that Veterans represent experienced and active wheelchair users (39,52,55,63,69,99). While these explanations are factually grounded, they nonetheless result in systematic underrepresentation of women, non-veteran users, and individuals with other disability profiles. Moreover, despite robotic wheelchairs frequently being framed as solutions for an "aging society," very few studies directly involved older adults over 65 years of age. This demographic concentration raises concerns regarding external validity and inclusivity, as mobility needs, risk perception, cognitive load, and technology acceptance patterns may differ substantially across gender and age groups. The resulting bias mirrors broader trends in biomedical and technological research, where middle-aged male participants are disproportionately represented, potentially limiting the generalizability and equity of innovation outcomes.

### 4.5 Engineering-Dominated Research Teams

Across the reviewed studies, research teams were overwhelmingly composed of engineers (~89%), with comparatively limited participation from social scientists, designers, disability scholars, or persons with disabilities themselves. This disciplinary concentration narrows development to performance optimization, overlooking social and lived realities. Such configurations tend to reinforce medical-technical models of assistive technology development rather than broader socio-technical approaches that expand user capabilities by accounting for autonomy, environmental context, and everyday practice. At the same time, very few studies explicitly documented persons with disabilities as sustained collaborators or co-design partners (i.e. the MEBot program (39,52,55,63,69), and Leaman, J. F., et al. (60)) highlighting a persistent distinction between consultation and genuine co-production and reflecting enduring power asymmetries in "designing for" rather than "designing with."

### 4.6 Limited User Influence on Core Design

Across the reviewed studies, user involvement was predominantly concentrated in testing and evaluation phases, with relatively few projects engaging end users during early requirement-definition stages prior to system design. In most cases, users were invited to assess performance, workload, usability, or satisfaction after key architectural decisions had already been made. While such feedback may contribute to parameter tuning or incremental refinements, it rarely influences the core design of the robotic wheelchair. Even at this stage, evaluation outcomes seldom translated into documented iterative redesign cycles affecting core system development, leaving user input largely confined to peripheral rather than foundational technological decisions. This pattern reflects a form of post hoc validation rather than genuine participatory development. It is not user-centered design, but user evaluation of predetermined designs, a structural limitation that constrains the transformative potential of user involvement in robotic wheelchair research.

### 4.7 Systemic Patterns of Validation: Disconnection from Real-World Contexts and Users

The limited influence of end users, already confined largely to evaluation phases, is further reinforced by embedded limitations in how robotic wheelchairs are tested and validated. Across the reviewed studies, testing was overwhelmingly conducted in controlled laboratory settings, whether simulating indoor or outdoor environments, typically structured around standardized obstacles and predefined navigation tasks. Moreover, test protocols were most often defined according to engineering standards or clinical expertise, with little evidence of user involvement in shaping testing scenarios.

While such settings prioritize technical reliability, safety benchmarking, and compliance with established standards, they provide limited insight into performance under real-world conditions characterized by environmental variability, social interaction, and unpredictable navigation demands. Similar concerns have been raised in the broader smart wheelchair and assistive technology literature, which notes that practical usability in everyday contexts remains underexplored (37,100). Users themselves emphasize the need for testing in cluttered homes, dynamic public spaces, and outdoor environments rather than exclusively controlled laboratory settings (100).

This emphasis on technical validation reinforces a development model in which performance metrics and standards compliance take precedence over participatory and contextual considerations. As a result, system safety and usability may be overestimated under idealized conditions, while the complexity of daily mobility environments, and the lived expectations, risks, and priorities of users, remain insufficiently captured (37,100).

### 4.8 Contextual Limitations and Global Inequality

A final systemic limitation concerns the geographic concentration of robotic wheelchair research. The vast majority of reviewed studies were conducted in high-income countries (HICs), while contributions from low- and middle-income countries (LMICs) were comparatively scarce and often centered on affordability adaptations rather than full technological development. This imbalance is particularly significant given that the majority of persons with disabilities, over 80% according to the WHO, live in LMICs, where poverty and limited access to services reinforce health inequities (101,102). Contextual conditions in many LMICs can differ substantially from those assumed in HIC-based development models, including financial constraints, infrastructure instability, limited technical and maintenance capacity, and challenging environmental conditions (7,98). When technologies are primarily designed for HICs and expected to transfer seamlessly to LMICs, they risk reduced functionality, limited sustainability, or even unintended harm (7). The geographic concentration of innovation therefore not only limits generalizability across infrastructural contexts but also reinforces global technological asymmetries, leaving the majority of potential users underrepresented in the design and validation of advanced robotic mobility systems.

Taken together, these findings reveal that the limited implementation of user-focused design methodologies in robotic wheelchair development is not a matter of isolated methodological oversight, but the result of interconnected systemic factors operating across the research lifecycle. From restricted inclusion and demographic bias to engineering-dominated teams, laboratory-centered validation, fragmented measurement practices, and geographic concentration in high-income settings, user perspectives are structurally constrained at multiple levels. Even when participation occurs, it is frequently positioned late in the process, limiting its capacity to shape core architectural decisions.

This misalignment has practical consequences. Evidence from assistive technology research consistently shows that technologies developed with sustained user involvement are more likely to demonstrate usability, accessibility, and acceptability (37,94). Conversely, insufficient engagement with end users has been associated with frustration, nonuse, and device abandonment (103,104). The gap between promising laboratory prototypes and sustained real-world adoption is therefore unlikely to be explained by user resistance to innovation. On the contrary, wheelchair users have expressed clear interest in advanced features such as autonomous navigation (105). The challenge lies not in lack of demand, but in the persistent structural misalignment between how technologies are developed and how they are expected to function in daily life.

## 5 CONCLUSIONS

Robotic wheelchairs represent a technologically promising response to persistent mobility challenges faced by persons with disabilities. Yet this review indicates that user-involvement design principles remain inconsistently and incompletely implemented in their development. Despite well-documented benefits of early and sustained end-user engagement in assistive technology design, including improved usability, accessibility, and acceptance (37,94), much robotic wheelchair research continues to prioritize technical advancement over meaningful participatory integration.

This narrative literature review shows a recurrent gap between technological sophistication and lived usability. This gap has tangible consequences: Across assistive technology domains, insufficient user involvement has been linked to dissatisfaction and device abandonment (103,104). In the case of advanced powered wheelchairs, several commercially released systems with innovative capabilities have failed to remain on the market (103), reflecting the broader translation challenge between laboratory innovation and sustainable real-world adoption. Notably, available evidence does not suggest a lack of user interest in advanced mobility technologies. On the contrary, potential users have expressed willingness to adopt features such as autonomous navigation when these align with their expectations and sense of control (105). The issue, therefore, is not user reluctance but structural misalignment in development processes.

Existing evidence clearly establishes that meaningful user engagement improves outcomes. Universal and participatory design approaches implemented from the idea-creation phase onward are associated with greater accessibility and usability (94), and insufficient attention to social and contextual factors has been linked to limited real-world deployment in human–robot interaction research more broadly (95). Involving users throughout the design lifecycle is not a “nice to have” enhancement but a foundational requirement for socially responsible and sustainable innovation.

While technology researchers and engineers play a central role in prioritizing participatory practices within their projects, framing limited user involvement solely as a failure of individual developers would be reductive. The evidence instead points to a constellation of institutional, regulatory, and organizational constraints that operate across the research and healthcare ecosystem. Regulatory and ethics procedures often require finalized protocols before early exploratory engagement can occur, effectively postponing user involvement until late-stage testing (98). Academic incentive structures continue to reward technical novelty and rapid publication over iterative, interdisciplinary collaboration, discouraging the slower processes required for genuine co-design (98,106–108). Within the healthcare ecosystem, wheelchair users are often among the least empowered stakeholders relative to institutional decision-makers, such as healthcare providers, insurers, manufacturers, and distributors, limiting their capacity to influence design and deployment decisions and posing a significant challenge for product adoption of durable medical and assistive devices (109). These dynamics do not excuse limited user integration, but they help explain why user-centred methodologies remain rare despite their recognized value.

### 5.1 Recommendations

Based on these findings, we propose the following:

A. **Integrate persons with disabilities as core members of development teams.** Participation should extend beyond consultation, with clearly defined roles and meaningful influence on requirement definition, design trade-offs, and iterative refinement. Tokenistic involvement must be avoided.

B. **Predefine study protocols and provide a clear rationale for sample size.** Sample size considerations are critical in both qualitative and quantitative research. Clearly defining the innovation goal and the metrics used to evaluate it helps ensure methodological coherence. Valuable lessons can be drawn from clinical research methodology in medicine and biomedical technology. While larger samples are desirable for confirmation and comparison, early-phase studies (e.g., Phase I) may appropriately focus on safety and preliminary effectiveness before progressing to more robust testing, provided that limitations are transparently acknowledged. International collaboration may facilitate the larger sample sizes required for confirmatory studies.

C. **Use at least one validated outcome instrument.** Incorporating standardized measures (e.g., SUS, NASA-TLX, QUEST, WST-Q, WHODAS) enhances comparability across studies and strengthens the evidentiary basis of user feedback.

D. **Document how user input informs design decisions.** Reporting should transparently indicate which suggestions were implemented, modified, or not adopted, and provide justification for those decisions.
E. **Complement laboratory testing with real-world validation.** When safety and feasibility allow, at least one evaluation phase should occur in realistic environments involving persons with disabilities.
F. **Plan and resource interdisciplinary collaboration.** If a fully interdisciplinary core team is not feasible, structured partnerships or advisory mechanisms should provide complementary expertise across technical, clinical, and social domains.
G. **Strengthen reporting standards in assistive technology research.** Journals should require rigorous and transparent documentation of participant characteristics, validated measures, testing environments, and evidence of iterative redesign when claims regarding autonomy, usability, or quality of life are made.

Ultimately, improving robotic wheelchair development is not solely a technical challenge but a systemic one. Without structural change, the field risks continuing to produce technically impressive systems that fail to achieve durable adoption. With it, robotic mobility technologies can more fully realize their potential to expand autonomy, participation, and quality of life.

## ACKNOWLEDGMENTS

This work was supported in part by the Swiss National Science Foundation (Ambizione award 186035; MRV).

# A APPENDICES

## A.1 Composition of Research Teams, Geographic Distribution, and Stakeholder Participation in the Reviewed Robotic Wheelchair Studies

| Ref | Multidisciplinarity (Team composition) | Countries (Centers/Users) | Codesign (PWD Author) | Other Stakeholder Involvement & Tasks |
|---|---|---|---|---|
| **(29)** | **Engineers (4).** | **China** | **None Reported** | **Experts: define requirements** |
| **(52)** | **Engineers with degrees in Rehabilitation Science (5), MD (1), and Physical Therapists (1).** | **USA** | **Yes: R. Cooper, B. Daveler** | **Physical Therapists (safety), university's machine shop staff (development) and clinical**<br><br>**coordinators (evaluation assistance)** |
| **(53)** | **Engineer** | **Japan** | **None Reported** | **Undergraduate Students: Assisted with the tests** |
| **(54)** | **Low: Engineers (4) and Clinical Psychologist (1).** | **Portugal** | **None Reported** | **Therapists, caregivers and psychologist: help during test for safety**<br><br>**Association Staff: ergonomic adaptations.** |
| **(32)** | **Engineers (4) and Product Designer (1).** | **USA / Users in 7 countries** | **None Reported** | **Researchers, Healthcare personnel, nonprofit**<br><br>**organizations, communities & Industry members: Users and context studies** |
| **(30)** | **Engineers (13), MDs (1), and Sociologists (1).** | **France and UK** | **None Reported** | **Robotics experts, occupational therapists, and specialists in rehabilitation: Define requirements (Regular roundtable sessions)** |
| **(62)** | **Engineers (4)** | **France** | **None Reported** | **Medical Equipment Company: Design and development** |
| **(66)** | **Engineers (3), Rehab Physicians (3), and Sociologists (1).** | **France** | **None Reported** | **Occupational Therapists: Evaluated the tests (to ensure safety standards)** |
| **(67)** | **Engineers (5), PRM Physicians (3), and Sociologists (1).** | **France and Canada** | **None Reported** | **Healthcare Professionals: Evaluated difficulty of the tests (not the technology)** |
| **(65)** | **Engineers (5)** | **France and Japan** | **None Reported** | **Occupational Therapists and specialists in rehabilitation: Expert consultation** |
| **(55)** | **Engineers with degrees in Rehabilitation Science (5), and Physical Therapists (1).** | **USA** | **Yes: R. Cooper, B. Daveler** | **Veterans Organizations** |
| **(56)** | **Engineers (5)** | **India** | **None Reported** | **None Reported** |
| **(57)** | **Engineers (2)** | **Sri Lanka** | **None Reported** | **Caretakers: "Control Experiments"** |
| **(58)** | **Engineers (3)** | **Philippines** | **None Reported** | **None Reported** |
| **(59)** | **Engineers (5)** | **Mexico** | **None Reported** | **None Reported** |
| **(60)** | **Engineers (4). And Astrophysics+ Engineer(1)** | **USA** | **Yes: J. Leaman** | **Undergraduate students : define requirements** |
| **(39)** | **Engineers (3), Engineers with degrees in Rehabilitation Science (2), MDs (1), and Physical Therapists (1).** | **USA** | **Yes: R. Cooper, B. Daveler** | **Physical therapist:**<br><br>**seat configuration**<br><br>**Spotters: safety during tests** |
| **(61)** | **Engineer (1)** | **Japan** | **None Reported** | **None Reported** |
| **(63)** | **Engineers with degrees in Rehabilitation Science (6)** | **USA** | **Yes: R. Cooper, B. Daveler** | **Licensed Professionals: Define requirements** |
| **(69)** | **Engineers with degrees in Rehabilitation Science (5), MDs (1), and Physical Therapists (1).** | **USA** | **Yes: R. Cooper, B. Daveler** | **Physical Therapists: Evaluation & configuration** |
| **(64)** | **Engineers (5) and Engineer-Physicist (1)** | **UK, USA, and France** | **None Reported** | **None Reported** |
| **(68)** | **Engineers (2)** | **China** | **None Reported** | **Experts & Guardians: define requirements.** |
| **(33)** | **Engineers (5)** | **Indonesia** | **None Reported** | **Caregivers: define requirements** |

**A.2 Overview of Methodological Frameworks, User-Reported Measures, Experimental Methods, and Testing Environments in Robotic Wheelchair Research**

| **Ref** | **General Methodology** | **User-Reported Data Collection Methods** | **Methods for Experimental Data Collection** | **Testing Environment** | **Standards and Protocols guiding the Test design** |
|---|---|---|---|---|---|
| (29) | User-Centered Design<br>Kano model | Questionnaires & interviews. | Physiological Measurements: EEG recording<br>Ergonomic simulation (CATIA) | Simulation:<br>Indoor (bedroom) and outdoor (parks). | Computer modeling and Ergonomic simulation (CATIA). |
| (52) | Heuristic Approach | None reported | Interaction and Performance Measurements: | Lab (Streets environment): Controlled curb | ISO 9241-11, ISO 7176-10, AASHTO Green Book, ADA |
| | | | efficiency/effectiveness metrics. | heights and sidewalk barriers. | accessibility guidelines. |
| (53) | None reported | None reported | Interaction and Performance Measurements: (speed, points and stability) | Lab/contest (Indoor and Outdoor environment): Competition circuit (Cybathlon), Ramps, rough terrain, stairs, slaloms. | Cybathlon rule book. |
| (54) | None reported | NASA-TLX questionnaire & customized satisfaction questionnaire | Quantitative performance Measurements | Real indoor office environment: Hallways, narrow doorways, office obstacles. | None reported |
| (32) | Participatory Design (PD) & Double Diamond (DD) | Customized questionnaires (open questions) and personal interviews. | None reported | No evaluation reported: Indoor Rooms (planed) | None reported |
| (30) | Codesign & Human centric design | Roundtable sessions<br>USE questionnaire; satisfaction written surveys. | Interaction and Performance Measurements: (driving, collisions, etc.) | Lab (Indoor environment): Elevators, slopes, corridors, and curb ramps. | Expert consultation (clinicians) |
| (62) | None reported | QUEST-like questionnaire. | Interaction and Performance Measurements: (driving, collisions, etc.) Random double-blind clinical trials; | Real world (Indoor): Hallway of the rehabilitation center during working hours with dynamic pedestrian traffic. | Expert consultation (medical staff) |
| (66) | None reported | NASA-TLX; USE; WST-Q; customized satisfaction assessments. | Interaction and Performance Measurements. Performance Assessment: WST & PIDA Single-blind randomized trials | Lab (Indoor environment): Standardized corridors and slopes (5°, 10°). | Lab/Clinical: Standardized reproduction circuits. Standard: WST and PIDA recommendations. & Expert consultation (350 PWC professional trainers) |
| (67) | None reported | NASA-TLX; USE; WST-Q (predictability). | Interaction and Performance Measurements. Performance Assessment: WST & PIDA Randomization of assisted vs. non-assisted modes | Lab (Indoor environment): Standardized narrow corridors and elevators. | PIDA, WST, WSP & Expert consultation (366 PWC professional trainers) |
| (65) | None reported | Satisfaction questionnaire | Interaction and Performance Measurements. (Time, collisions, obstacle detection, etc) | Lab (Indoor environment): Circuit with clutters, slaloms, and elevators. | None reported |
| (55) | User-Centered Design (UCD) | NASA-TLX; System Usability Scale (SUS). | Interaction and Performance Measurements. | Lab (Outdoor environment): environmental barriers, curb ramps. | ADA ramp safety guidelines and AASHTO. |
| (56) | User-Centered Design | None reported | Interaction and Performance Measurements: (speed, transition times, etc) | Lab (Indoor environment): Internal 15-meter road testing. | None reported |
| (57) | None reported | User satisfaction rating: OK - NotOK). | Interaction and Performance Measurements: Intelligent vs Manual wheelchair comparison | Lab (Indoor environment): Room with obstacles. | None reported |
| (58) | None reported | acceptance questionnaire based on the Likert Scale | None reported | Lab: flat and inclined paths | None reported |
| (59) | None reported | None reported | Interaction and Performance Measurements: Accuracy tests for head motion, voice, and muscle modes. | Lab (Indoor environment): Laboratory hallways and testing benches. | None reported |
| (60) | Participatory Design | Q&A sessions | Virtual simulations | Virtual simulation (Indoor & Outdoor): Bedroom, kitchen, living room, and bus stop. | None reported |
| (39) | User-Centered Design (UCD) | Standardized surveys: QUEST 2.0 (Likert scale), SUS questionnaire | Interaction and Performance Measurements: (Quantitative driving metrics). | Lab (Outdoor environment): Ramps (10°, 8°), cross slopes, and potholes. | ISO 7176-2, ADA compliance. Literature reporting user interviews (Veterans) |
| (61) | None reported | None reported | Interaction and Performance Measurements: (Simulation (Open Dynamics Engine) vs. Reality)) | Simulation & Lab (Outdoor environment): Oblique traversal patterns for steps. | None reported |
| (63) | Participatory Design (PD) | Questionnaires& focus group discussions, Likelihood-of-use survey. | None reported | Lab: Videos and demonstrations. (Intended for outdoor use) | User consultation on driving conditions. |
| (69) | None reported | Focus groups, NASA-TLX questionnaire, satisfaction surveys; interviews. | Interaction and Performance Measurements Powered Mobility Clinical Driving Assessment (PMCDA) | Lab (Outdoor environment): Ramps, curbs, cross-slopes. | ISO 9241-11, American with Disabilities Act's standards for accessible design (ADA). |
| (64) | None reported | IBM Computer Usability Satisfaction Questionnaire (CSUQ) | Wheelchair Skills Test Interaction and Performance Measurements:(distance travelled; time; clearance, etc) | Lab (Outdoor environment): course with cardboard box obstacles and reversing into an elevator. | WST key navigation tasks |
| (68) | None reported | Interview survey and questionnaire. | Technical testing with dummy (not real users). | Simulation | GB/T13800-92 |
| (33) | User-Centered Design (UCD) AHOQ (Axiomatic House of Quality), | ENASE-based questionnaires; Open-ended questionnaire; Likert ratings. | Finite Element Analysis (FEA) - (Simulation - no users) | Simulation: Intended for outdoor and slope navigation | ISO 7176-5 |

### A.3 Methods and Instruments Used to Collect User Feedback in Reviewed Robotic Wheelchair Studies

| Category | Method / Tool (Studies) | Frequency Among Included Studies | Brief Description |
|---|---|---|---|
| Qualitative Methods | Interviews (29,32,68,69) | 4 | Structured or semi-structured verbal data collection. |
| | Focus Groups (63,69) | 2 | Group discussions to explore shared experiences and needs. |
| | Roundtable Sessions (30) | 1 | Structured multi-stakeholder discussion format. |
| | Q&A Sessions (60) | 1 | Interactive design feedback sessions. |
| | Open-Ended Questionnaires (32,33) | 2 | Free-text surveys for requirement elicitation. |
| Customized Questionnaires | Customized Questionnaires (32,54,57,58,63,65,66,69) | 8 | Non-standardized satisfaction or usability surveys. |
| Standardized Usability & Satisfaction Scales | SUS (39,55) | 2 | Standardized system usability questionnaire. |
| | USE (30,66,67) | 3 | Measures usefulness, satisfaction, and ease of use. |
| | QUEST / QUEST 2.0 (39,62) | 2 | Assistive technology satisfaction scale. |
| | WST-Q (66,67) | 2 | Self-reported wheelchair skill assessment. |
| | CSUQ (64) | 1 | IBM usability satisfaction questionnaire. |
| | ENASE-based Questionnaires (33) | 1 | Structured satisfaction evaluation framework. |
| Workload & Cognitive Load Measures | NASA-TLX (54,55,66,67,69) | 5 | Standardized perceived workload scale. |